\documentclass{ws-mpla}
\usepackage{cite}
\usepackage{graphicx}
\usepackage{lipsum}
\usepackage{bm}
\usepackage{mathrsfs}  
\usepackage{slashed}
\usepackage{subfigure}
\usepackage{feynmp}
\usepackage{graphicx}
\usepackage{amsfonts}
\usepackage{color}
\usepackage{cancel}
\usepackage{amsmath}
\usepackage{makecell}

\def\Tr{\mbox{Tr}\,}

\def\hbar{\hspace{0pt}\raisebox{1pt}{$-$} \hspace{-7pt} h}

\def\5{\overline 5}

\definecolor{JJ}{RGB}{0,144,255}

\newcommand{\be}{\begin{equation}}
\newcommand{\ee}{\end{equation}}
\newcommand{\bea}{\begin{eqnarray}}
\newcommand{\eea}{\end{eqnarray}}

\newcommand{\ba}{\begin{eqnarray}}
\newcommand{\ea}{\end{eqnarray}}

\definecolor{rossoCP3}{cmyk}{0,.88,.77,.40}
\definecolor{graa}{rgb}{0.8,0.8,0.8}
\definecolor{blaa}{rgb}{0.2,0.2,0.6}

\begin{document}

\markboth{Emiliano Molinaro and Natascia Vignaroli}
{Diphoton Resonances at the LHC}

%%%%%%%%%%%%%%%%%%%%% Publisher's Area please ignore %%%%%%%%%%%%%%
\catchline{}{}{}{}{}
%%%%%%%%%%%%%%%%%%%%%%%%%%%%%%%%%%%%%%%%%%%%%%%%%%%%%%%%%%%%%%%%%%%

\title{ Diphoton Resonances at the LHC}

\author{\footnotesize Emiliano Molinaro\footnote{molinaro@cp3-origins.net} \hspace{0.15cm} and \hspace{0.15cm}  Natascia Vignaroli\footnote{vignaroli@cp3-origins.net}}
\address{CP$^{3}$-Origins, University of Southern Denmark\\
Campusvej 55, DK-5230 Odense M,
Denmark
}

\maketitle

%\pub{Received (Day Month Year)}{Revised (Day Month Year)}

\begin{abstract}
We review the current status of searches for new physics beyond the Standard Model in the diphoton channel at the LHC and estimate the reach with future collected data. We perform a model independent analysis based on an effective field theory approach and different production mechanisms. As an illustrative example,  we apply our results to a scenario of minimal composite dynamics.

%\keywords{Keyword1; keyword2; keyword3.}
\end{abstract}

%\ccode{PACS Nos.: include PACS Nos.}

\section{Introduction}

Searches for resonances in diphoton events are among the highest-priority tasks of the experimental program ongoing at the LHC. Indeed, the analysis of the diphoton channel led to the discovery of the Higgs boson back in 2012 
\cite{Aad:2012tfa,Chatrchyan:2012xdj}. On general grounds, this is one of the cleanest signal at the LHC and with a relatively simple search strategy it is possible to probe via this channel different scenarios of new physics beyond the Standard Model (SM). 

The latest experimental  analyses on diphoton resonances from ATLAS \cite{ATLAS:2016eeo} and CMS \cite{Khachatryan:2016yec} at a beam colliding energy $\sqrt{s}=13$ TeV (LHC-13) have considered a total integrated luminosity of 15.4 fb$^{-1}$ and 12.9 fb$^{-1}$, respectively. 
In this brief review we focus on these  results in order to set model independent limits on new physics scenarios which feature a spin-0 state $X$ coupled to photons, with a mass in the TeV range. We also  provide an estimate of the future reach at LHC-13 with different luminosities. We adopt an effective description of the phenomenology of the particle $X$ and focus on 
two different production mechanisms, which apply to a vast category of models, that is: $i)$ the new particle $X$ is mostly generated via photon fusion and $ii)$ $X$ is directly coupled to the top quark, which induces dominant production through gluon fusion, in analogy with the SM Higgs boson. 

Models of new physics beyond the SM to which our analysis applies  
include, for example, theories of minimal composite dynamics \cite{DiVecchia:1980xq, Molinaro:2015cwg, Molinaro:2016oix, Tandean:1995ci,Gripaios:2009pe, Galloway:2010bp, Bellazzini:2015nxw, Xue:2016txt}, where $X$ is a composite  pseudoscalar state, analogous to the $\eta$ or the $\eta^\prime$ in QCD, and theories with axion-like particles  (see, e.g.,\cite{Mimasu:2014nea} and references therein). Similar constraints in the diphoton channel apply to scenarios which predict spin-2 resonances coupled to photons, such as the graviton in Randall-Sundrum models \cite{Randall:1999ee} or spin-0 radion/dilaton particles \cite{Goldberger:1999uk,Sundrum:2003yt, Coradeschi:2013gda, Bellazzini:2013fga, Megias:2014iwa,Vecchi:2010gj,Ahmed:2015uqt}. Recent studies of the diphoton channel have also considered models with vector-like fermion dark matter \cite{Gopalakrishna:2017zku}, two Higgs doublets \cite{Bian:2017jpt}, $R-$axion \cite{Bellazzini:2017neg}, next-to-minimal supersymmetric SM \cite{Cao:2016uwt} and more involved topologies leading to a diphoton signature \cite{Allanach:2017qbs}.
As an illustrative example, we use our results to extract information on the fundamental parameters of a minimal composite theory of electroweak (EW) symmetry breaking, where the  particle $X$ is identified with a composite  pseudoscalar state associated with a global anomalous $U(1)$ symmetry. It interacts with EW gauge bosons via anomalous couplings originating from the topological sector of the theory, that is the gauged Wess-Zumino-Witten (WZW) action \cite{Witten:1983tx,Wess:1971yu,Kaymakcalan:1984bz,Kaymakcalan:1983qq,Schechter:1986vs}.

The paper is structured as follows:  
in section \ref{sec:xsec} we first introduce our effective description of the diphoton resonance and then we present the results on constraints and reach of the diphoton channel under the hypotheses of  photon fusion and top-mediated gluon fusion production mechanisms. In section \ref{sec:model} we discuss a technicolor-like theory and apply our results in order to constraints its fundamental parameters at the LHC-13. Finally, we summarize our results in section \ref{summary}.

\section{Effective description }\label{sec:xsec}

We use an effective approach to derive the LHC-13 reach on diphoton resonances. 
We consider a pseudoscalar boson $X$ and two scenarios where the dominant pseudoscalar production mechanisms are either photon fusion or top-mediated gluon fusion. New physics effects are encoded in the effective Lagrangian: 
\begin{eqnarray}\label{eq:L-eff}
\begin{split}
\mathcal{L}_{\rm eff}=& -i y_t \frac{m_t}{v} X\, \bar{t} \gamma_5 t - \frac{c_{GG}}{8v}  X\, \Tr\left[ G^{\mu\nu}\tilde{G}_{\mu\nu}\right] \\
& - \frac{c_{AA}}{8v}  X\, A^{\mu\nu}\tilde{A}_{\mu\nu} - \frac{c_{AZ}}{4v}  X\, A^{\mu\nu}\tilde{Z}_{\mu\nu}- \frac{c_{WW}}{4v}  X\, W^{+ \mu\nu}\tilde{W}^{-}_{\mu\nu} - \frac{c_{ZZ}}{8v}  X\, Z^{\mu\nu}\tilde{Z}_{\mu\nu}\,,
\end{split}
\end{eqnarray}
where $v=246$ GeV is the EW scale and $\tilde{V}^{\mu\nu}=\epsilon^{\mu\nu\rho\sigma}V_{\rho\sigma}$, for $V=G,A,Z,W^\pm$. 
Notice that the scale of new physics $\Lambda_{\rm NP}$ is absorbed into the definition of the Wilson coefficients
\begin{equation}\label{eq:c-lambda}
 c_{VV} \approx \frac{1}{4 \pi}\frac{v}{\Lambda_{\rm NP}} \ .
 \end{equation}
The latter can be either radiatively generated or can arise from some non-perturbative dynamics. They thus incorporate new physics effects and also the computable SM contributions coming from the renormalizable interactions of the pseudoscalar $X$ with SM particles. 

In our effective description we also include the possibility of a direct coupling of $X$ to the top quark, that is $y_t \neq 0$ in Eq. (\ref{eq:L-eff}).
 We consider that the top radiative contribution completely generates the $X$ effective coupling to the gluons, $c_{GG}$, which reads 
\begin{equation}
c_{GG}= y_t \frac{\alpha_S}{2 \pi}F\left(\frac{m_X^2}{4\,m_t^2}\right) \ ,\label{cGG}
\end{equation}
where $\alpha_S$ is the strong coupling constant, $m_t$ is the top mass and $F(x)=-\frac{1}{4x}\left(\ln\frac{\sqrt x+ \sqrt{x-1}}{\sqrt x- \sqrt{x-1}}- i\pi\right)^2$ for $x>1$, see \cite{Steinberger:1949wx}. 
The top-loop contribution is included also in the other $c_{VV}$ couplings. However, we will take them as free parameters in order to analyze new physics effects.

Our effective description includes up to dim-5 operators. Corrections from dim-6 operators can be estimated to be of the order 
$\left( m_X/4 \pi \Lambda_{NP}\right)^2 \approx \left( c_{VV}  m_X/v \right)^2$.
These corrections become sizable in the region of the parameter space corresponding to large values of the effective couplings and the $X$ mass, $m_X$. For example, using the previous estimate, we expect corrections $\gtrsim$ 50\% for $m_X \gtrsim$ 4 TeV and $c_{VV}\gtrsim 0.045$.

In our analysis we will focus on two cases, one in which $X$ is not coupled to the top, $y_t =0$, and one in which the coupling to the top is of order $y_t \simeq 1$. %, still within the perturbative regime. 
In the first case the new state $X$ is mostly produced via photon fusion, while in the second case we will have  top-mediated gluon fusion production. Notice that for $y_t \simeq 1$, still within the perturbative regime, the total width of $X$ is naturally enhanced by the tree-level decay rate to tops and we do not expect the produced resonance to be narrow.  
  
\begin{figure*}[t!]
\begin{center}
\includegraphics[width=0.48\textwidth]{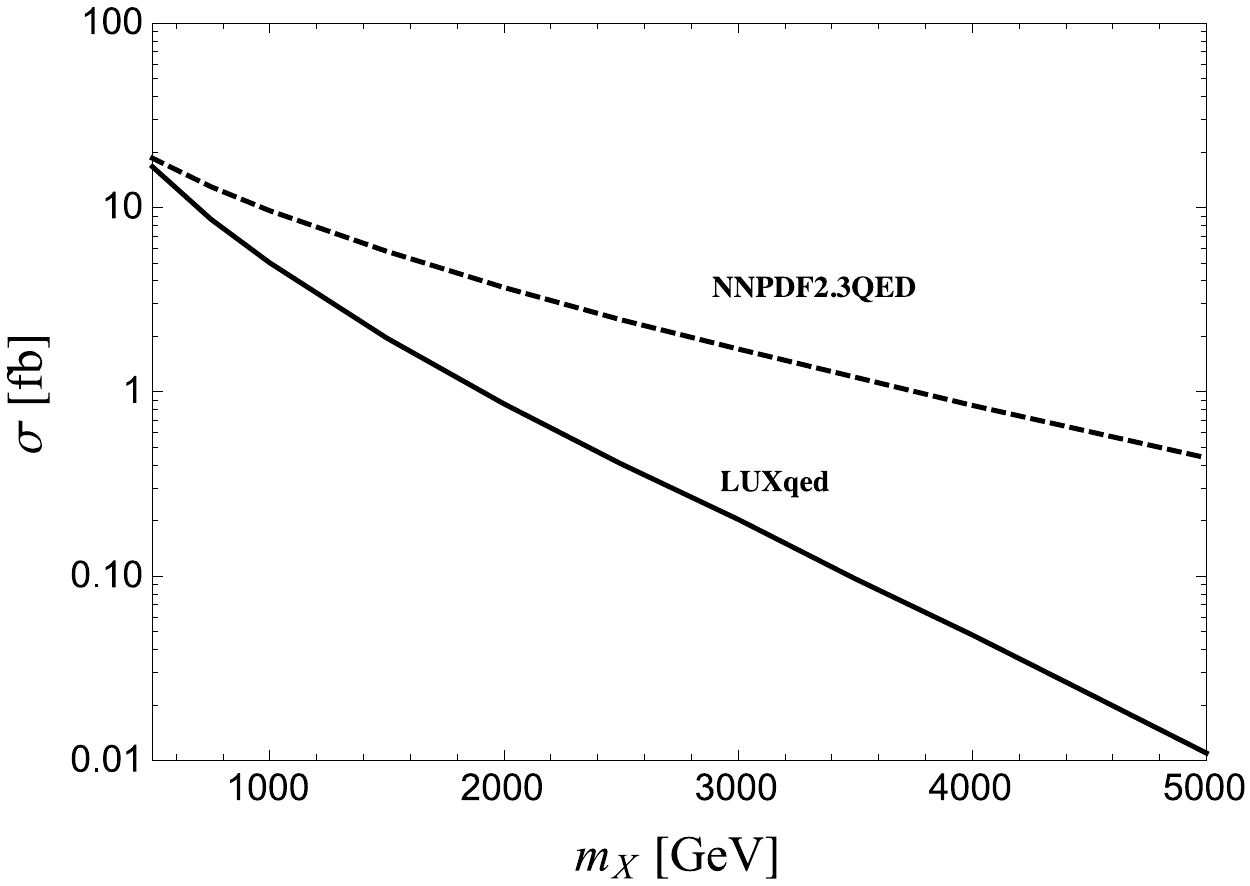}
\includegraphics[width=0.49\textwidth]{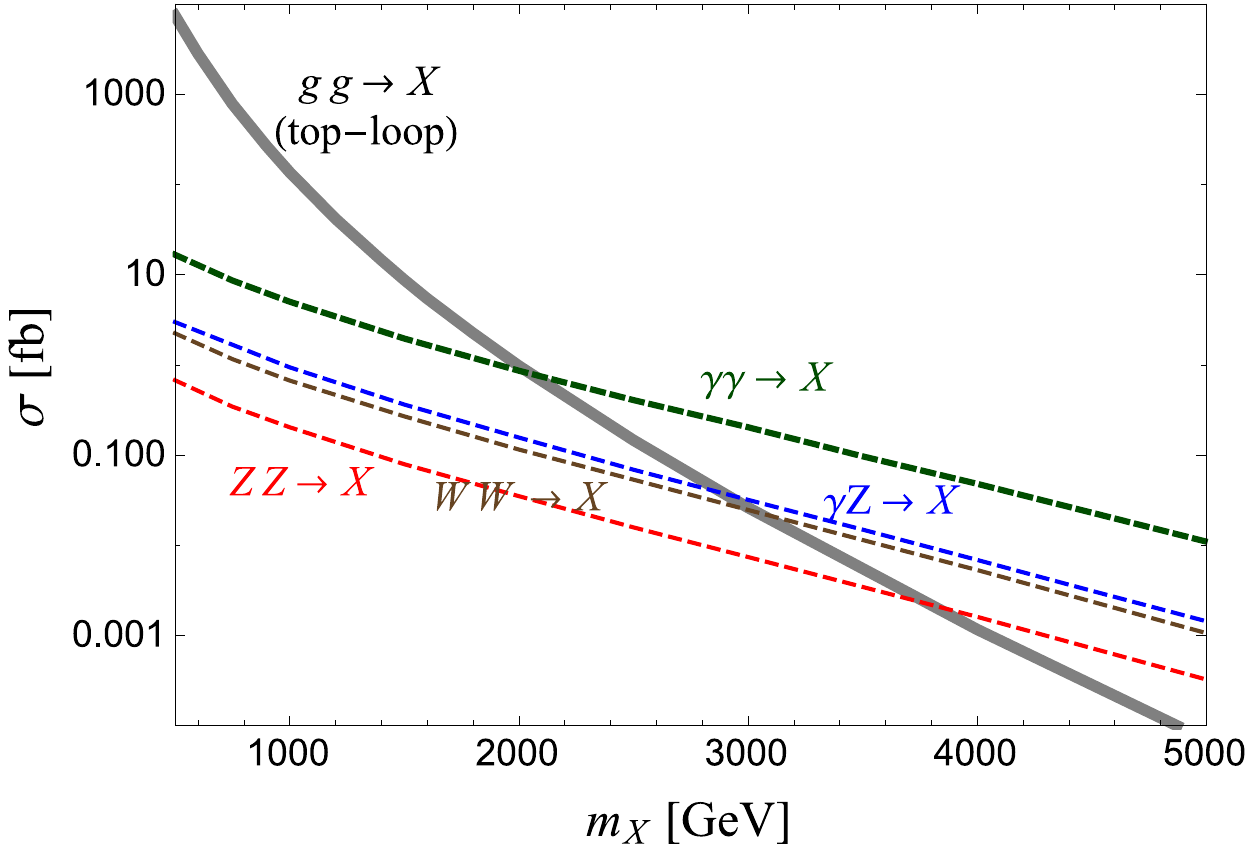} \,~~~~~~~~
\caption{\textbf{Left panel:} $X$ production cross section via photon fusion at LHC-13 using two different sets of photon PDF: LUXqed \cite{Manohar:2016nzj} and NNPDF2.3QED \cite{Ball:2013hta}. We set $c_{AA}=0.05$. \textbf{Right panel:} LHC-13 cross sections as a function of $m_X$ for different production channels: $\gamma\gamma$, $\gamma Z$, $WW$ and $ZZ$ fusion (for $c_{AA}=c_{AZ}=c_{WW}=c_{ZZ}=0.05$) and top-mediated gluon fusion (for $y_t=1$), see Eq.~(\ref{eq:L-eff}). The gluon fusion cross section is calculated at NLO in QCD. Here we use the LUXqed PDF.}
\label{fig:xsec}
\end{center}
\end{figure*}

 As a starting point of our analysis, we plot in the right panel of Fig.~\ref{fig:xsec} the production cross sections at LHC-13 for various mechanisms, stemming from our effective action, as a function of $m_X$. For illustration, we assume $y_t=1$ for the gluon fusion production and $c_{VV}=0.05$ for the other mechanisms. 
 
The photon fusion production receives three contributions: the dominant one comes from inelastic/incoherent scattering, whereas two subleading contributions arise from the semi-elastic and the elastic scattering processes \cite{Fichet:2013gsa,Fichet:2014uka,Harland-Lang:2016kog}, where either one of or both the colliding protons remain intact. We calculate the production cross section at leading order (LO) with  MadGraph5\_aMC@NLO \cite{Alwall:2014hca}. 
  Both the elastic and inelastic  photon emission are included in the  $\gamma Z$ production cross section.
The latter and the remaining vector boson fusion contributions from $ZZ$ and $WW$ are also calculated with MadGraph5 and are subdominant  when the effective couplings $c_{VV}$ are of the same size. This is typically realized when they are generated by a common source of new physics.
Photon fusion and $\gamma Z$ cross sections are computed using the new set of photon parton distribution function (PDF), LUXqed \cite{Manohar:2016nzj}. LUXqed has significantly reduced the uncertainty on the photon fusion cross section compared to previous PDF sets as NNPDF2.3QED \cite{Ball:2013hta},  MRST2004QED \cite{Martin:2004dh} and the recent CT14QED \cite{Schmidt:2015zda}. For the NNPDF2.3QED set, the quoted uncertainty is typically of the order of 50\% \cite{Ball:2013hta} and even bigger for large momentum fraction  $x$, $x\gtrsim 0.1$ ($m_X \gtrsim 1.3$ TeV at LHC-13). The uncertainty for LUXqed is much smaller, of the order of 1-2\% over a large range of $x$ values \cite{Manohar:2016nzj}. %(cp. also with \cite{Harland-Lang:2016kog}). 
In the left panel of Fig.~\ref{fig:xsec} we compare the cross section for the photon fusion production of the pseudoscalar obtained with the LUXqed and the NNPDF2.3QED set, fixing $c_{AA}=0.05$. LUXqed predicts a cross section which is order of magnitudes smaller than that coming from NNPDF2.3QED set. The latter is the one used in our previous study \cite{Molinaro:2016oix}. We thus expect a significative decrease of the reach for heavier pseudoscalars, $m_X \gtrsim 3$ TeV, compared to the one estimated in \cite{Molinaro:2016oix}.\footnote{Indeed, in \cite{Molinaro:2016oix} we highlighted the importance of a precise determination of the photon PDF.} 

The cross section for gluon fusion is calculated at next-to-leading order (NLO) in QCD. K-factor corrections range from 2.1 to 1.8 for 0.5 TeV $\lesssim m_X \lesssim$ 2 TeV. We estimate them  using the model in \cite{Demartin:2014fia}. We note that the gluon fusion cross section drops quickly with $m_X$. This is a combined effect of the top-loop function, which goes to zero in the limit $m_X/m_t \to \infty$, and the scaling of the gluon PDF at large momentum fraction. A minor effect is given by the running of the strong coupling constant $\alpha_S$. Compared to gluon fusion, the cross sections for weak gauge boson fusion, especially for photon fusion, scale much more gently with $m_X$. As a consequence, we expect  a wider reach on the pseudoscalar mass for a diphoton resonance produced via photon fusion.

 \subsection{Photon fusion production}\label{subsec:photon-fusion}

We derive the exclusion regions and the expected reach on the Wilson coefficient $c_{AA}$ in Eq.~\eqref{eq:L-eff}  and the $X$ mass, under the minimal assumption that the new particle couples dominantly to photons.  
In this case the production cross section at LO in the narrow width approximation is given by
\begin{equation}\label{eq:xsec}
\sigma(p p \to X \to \gamma\gamma) = \frac{8 \pi^2 \,\Gamma(X \to \gamma\,\gamma)}{m_X}\frac{d \mathcal{L^{\gamma\gamma}}}{d m^2_X} \, \text{BR}(X\to \gamma\,\gamma) \ ,
\end{equation}
where $d \mathcal{L^{\gamma\gamma}}/d m^2_X$ is  the photon luminosity function, which can be deduced from the corresponding  cross section reported in Fig.~\ref{fig:xsec}. Notice that the production cross section depends quadratically on $c_{AA}$.

\begin{figure}[]
\begin{center}
\includegraphics[width=0.6\textwidth]{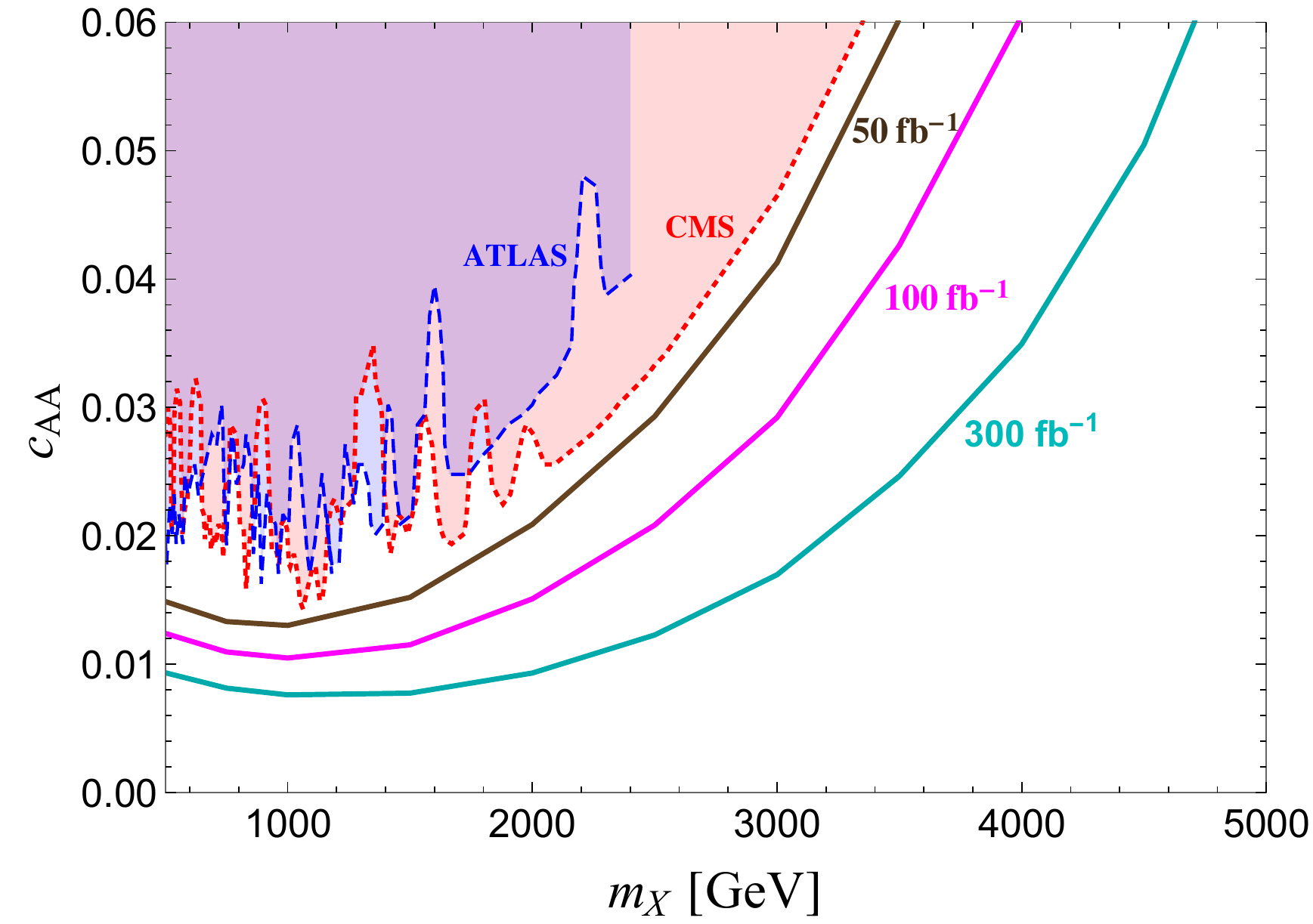} 
\caption{LHC-13 current constraints (dashed and dotted lines) and expected future reach of the diphoton channel in the plane ($m_X$, $c_{AA}$). We assume that $X$ is totally produced by photon fusion and $\text{BR}(X\to \gamma\gamma)=1$. The continuous curves show the 2$\sigma$ reach for different integrated luminosities.
 }
\label{fig:aaF}
\end{center}
\end{figure}

We report in Fig.~\ref{fig:aaF} the current experimental limits and the future reach on $c_{AA}$ as a function of the $m_X$, taking $\text{BR}(X\to \gamma\gamma)=1$. Results corresponding to a different branching ratio can be easily derived by properly rescaling the lines reported in the figure.\footnote{As mentioned in the previous section, one has to keep in mind that the effective coupling $c_{AA}$ is expected to receive sizable corrections from higher order operators in the effective Lagrangian for $c_{AA}^2 m_X^2 /v^2\approx\mathcal{O}(1)$.} 
We indicate in the figure the present   95\% C.L. limits at LHC-13 on narrow diphoton resonances. 
They correspond to the
ATLAS analysis with 15.4 fb$^{-1}$ \cite{ATLAS:2016eeo} (blue dashed curve) and the CMS analysis  with 12.9 fb$^{-1}$ \cite{Khachatryan:2016yec} (red dotted curve). We find that these limits are much stronger than
the ones obtained in the LHC-8 analysis from ATLAS \cite{Aad:2015mna} with 20.3 fb$^{-1}$ and CMS \cite{Khachatryan:2015qba} with 19.7   fb$^{-1}$.

To derive the LHC-13 reach of the diphoton channel we use the analytic estimate of the background given in the ATLAS analysis   \cite{ATLAS:2016eeo}, which yields the following best-fit function for the number of background events:
\begin{equation}\label{eq:bckg}
B(x, L) = \frac{L}{15.4\,\text{fb}^{-1}}\left(1-x^{1/3}\right)^{9.9} x^{-2.6}\,,
\end{equation}
where $x=m_{\gamma\gamma}/\sqrt{s}$, $m_{\gamma\gamma}$ is the diphoton invariant mass and $L$ is the total integrated luminosity. 
Notice that the background function in the new analysis \cite{ATLAS:2016eeo} has slightly changed from the one used in the ATLAS analysis given in \cite{Aaboud:2016tru} with 3.2 fb$^{-1}$.

In our determination of the reach we apply the selection criteria reported in \cite{ATLAS:2016eeo}, that is
\begin{equation}\label{eq:cuts}
E^{\gamma_1}_T > 0.4 \, m_{\gamma \gamma} \ , \quad  E^{\gamma_2}_T > 0.3 \,  m_{\gamma \gamma} \ , \quad |\eta_{\gamma_{1,2}}|<2.37 \quad ( |\eta_{\gamma_{1,2}}| \notin [1.37,1.52]) \ ,
\end{equation}
where $E_T^{\gamma_{1,2}}$ and $\eta_{\gamma_{1,2}}$ denote, respectively, the transverse energies and pseudorapidities of the two leading photons.
Furthermore, we assume a 95\% efficiency for the photon identification and we apply the isolation cut
\begin{equation}\label{eq:iso}
E^{\rm iso}_T < 0.05\, E^{\gamma_{1,2} }_T + 6 \, \text{GeV} \ ,
\end{equation}
where $E^{\rm iso}_T$  is defined as the transverse energy of the vector sum of all stable particles (except muons and neutrinos)  found within a cone $\Delta R =\sqrt{(\Delta{\eta})^2+(\Delta{\phi})^2 }\leq 0.4$, with $\Delta{\eta}$ ($\Delta{\phi}$) the pseudorapidity (azimuthal angle) separation.

We simulate the signal with  MadGraph5\_aMC@NLO \cite{Alwall:2014hca}, passing the events to PYTHIA 6.4 \cite{Sjostrand:2006za} for showering and hadronization and to DELPHES 3 \cite{deFavereau:2013fsa} to mimic detector effects. The resulting signal acceptances vary from 0.47 for $m_X=$ 0.5 TeV to 0.56 for $m_X=$ 5 TeV. 
The reach of the diphoton channel  is estimated by taking a sensitivity $S/\sqrt{S+B}=2$, where $S$ ($B$) denotes the number of signal (background) events passing the selection at a given integrated luminosity. 
The corresponding 95\% C.L. limits at LHC-13 are reported in Fig.~\ref{fig:aaF} for three choices of the integrated luminosity and apply to any new physics scenario with a new particle $X$ entirely produced via photon fusion, with $\text{BR}(X\to \gamma\gamma)=1$. Taking for example $m_X=1$ TeV, we obtain from current data that
 $c_{AA}\geq 0.018$ is excluded at 95\% C.L., while with future data LHC-13 will be able to test values of $c_{AA}\gtrsim 0.013 ~(0.010)~[0.0077]$ with a luminosity of 50 (100) [300] fb$^{-1}$.

 \subsection{Top-mediated gluon fusion production}\label{sec:GF}

 \begin{figure}[t!]
\begin{center}
\includegraphics[width=0.6\textwidth]{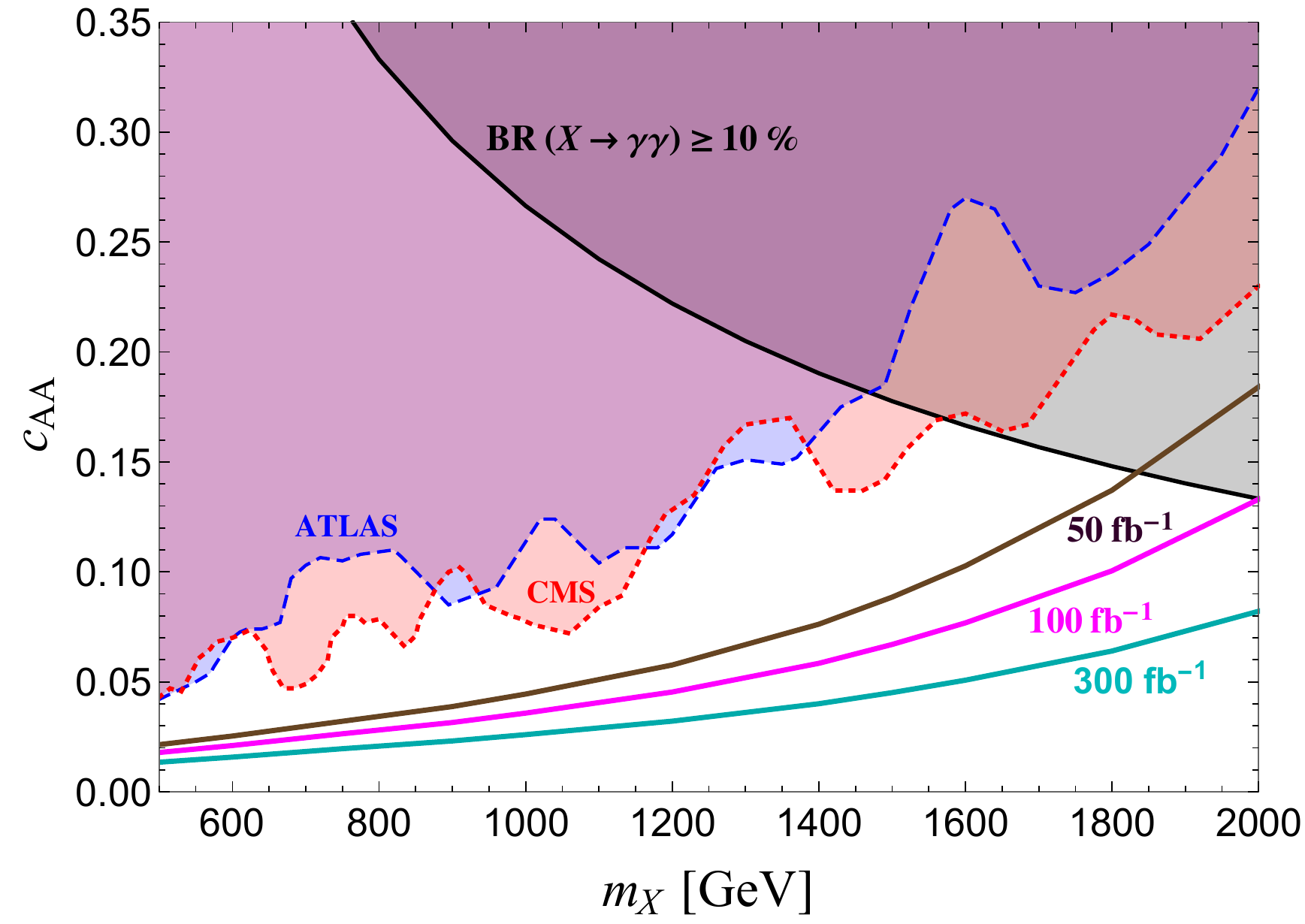} \,~~~~~~~~
\caption{
As in Fig. \ref{fig:aaF}, under the assumptions that $X$ is produced via top-mediated gluon fusion and has a total decay width $\Gamma_{\rm tot}(X) \simeq \Gamma(X \to t\bar{t})$.
The dark shaded region corresponding to ${\rm BR}(X\to\gamma\gamma)\geq 10\%$ indicates the parameter space where we expect sizable corrections to our predictions (see the text for details).}
\label{fig:ggF}
\end{center}
\end{figure}

 We now consider scenarios in which the pseudoscalar $X$ is directly coupled to the top via the renormalizable interaction given in Eq. \eqref{eq:L-eff}, that is we set $y_t\neq 0$.
  In this case $X$ dominantly decays at tree-level into a $t \bar{t}$ pair and its production at the LHC  proceeds via  top-mediated gluon fusion, with a signal cross section  given by
\begin{eqnarray}\label{eq:xsecgg}
\begin{split}
\sigma(p p \to X \to \gamma\gamma) & =  \frac{ \pi^2 \Gamma(X \to g\,g)}{8 \,m_X}\frac{d \mathcal{L}^{g g}}{d m^2_X} \times \text{BR}(X\to \gamma\,\gamma) \\
&\simeq  c^2_{AA} \frac{\alpha^2_S |F(m^2_X/4 m^2_t)|^2}{6144 \, \pi}\frac{m^4_X}{m^2_t v^2}\frac{d \mathcal{L}^{g g}}{d m^2_X}
\end{split}
\end{eqnarray}
where ${d \mathcal{L}^{g g}}/{d m^2_X}$ is the gluon luminosity function, which can be extracted from Fig.~\ref{fig:xsec} and $F(x)$ is the loop function defined below Eq.~\eqref{cGG}. Notice that  the production cross section is independent of $y_t$.
 
We repeat the analysis of the exclusion limits and the reach of the diphoton channel applied in the previous case and present our results in Fig.~\ref{fig:ggF}.  The production cross section  includes the NLO K-factor corrections, as discussed in section~\ref{sec:xsec}. We obtain in this case signal acceptances from 0.43 at $m_X=0.5$ TeV to 0.46 at $m_X=2$ TeV.   We report the current limits from ATLAS with 15.4 fb$^{-1}$ \cite{ATLAS:2016eeo} (blue dashed curve) and CMS with 12.9 fb$^{-1}$  \cite{Khachatryan:2016yec} (red dotted curve). 
 The continuous lines represent the 2$\sigma$ reach for different integrated luminosities. 
 The gray-shaded area in the plot corresponds to  $\text{BR}(X \to \gamma\gamma)\geq 10\%$, calculated for $y_t=1$.  In this region of the parameter space,  we expect a non-negligible contribution to the production cross section from the photon fusion mechanism and sizable corrections to the approximated expression in Eq.~\eqref{eq:xsecgg}. We obtain that values of $c_{AA}\geq 0.077$ are excluded at 95\% C.L. for $m_X=1$ TeV, while for 
 the same mass it is possible to probe the effective coupling $c_{AA}$ up to 0.045 (0.036) [0.027] with a luminosity of 50 (100) [300] fb$^{-1}$.
 
 As expected, because of  the different scaling of the photon and gluon PDFs with $m_X$, the LHC reach of the diphoton channel for a pseudoscalar $X$  dominantly produced via photon fusion is wider, extending to smaller effective couplings and to a larger mass range, compared to that of a $X$ which is mainly produced via gluon fusion.  
 
\section{LHC-13 tests of composite diphoton resonances}\label{sec:model}

We focus  on a minimal framework of composite dynamics which naturally features a new pseudoscalar particle in the TeV range, that can be produced at the LHC and decays in the diphoton channel. 
We consider a technicolor-like theory with 4 Weyl fermions $U_{L,R}$ and $D_{L,R}$ -- dubbed techniquarks -- in a complex representation $R$ of a new strong gauge group $SU(N_T) $, which do not carry SM color charge.
In the absence of EW interactions, the theory preserves a $SU(2)_L\times SU(2)_R$  global chiral symmetry which is dynamically broken around  $\Lambda_T \gtrsim 1$ TeV to the  custodial group $SU(2)_V$ by the techniquark condensate $\langle 0| \overline{U}_L U_R +\overline{D}_L D_R + \text{h.c.}|0\rangle \neq 0$. 

The three Goldstone bosons, the technipions, which arise from the breaking of the axial-vector symmetry are composite pseudoscalar fields made up of the techniquarks and their antiparticles. When the $SU(2)_W\times U(1)_Y$ gauge interactions are switched on, the technipions become the  longitudinal polarizations of the weak gauge bosons.
The left-handed techniquarks are combined in a doublet of $SU(2)_W$,  $Q_L\equiv(U_L, D_L)$, while the right-handed fields $U_R$, $D_R$ are weak isosinglets.
In order to cancel Witten  \cite{Witten:1982fp} and gauge anomalies, new fermions $E_{L,R}$ and $N_{L,R}$ are introduced, which are singlets of $SU(N_T)$. The left-handed chiral fields are also embedded in a $SU(2)_W$ doublet, $L_L \equiv(N_L, E_L)$. We consider the following  hypercharge assignments:
\begin{align}
\begin{split}\label{HYtechniquarks}
	& Y(Q_L) \,=\, \frac{y}{2}\,, \quad\quad Y(U_R/D_R)\,=\,\frac{y\pm1}{2}\,, \\ 
	&  Y(L_L)\,=\, -d(R)\,\frac y 2\,\quad\quad Y(N_R/E_R)\,=\,\frac{-d(R)\,y\pm 1}{2}  \, ,
	\end{split}
\end{align}
where $d(R)$ represents the dimension of the techniquark representation and $y$ is a real parameter. The electric charge is given by $Q=T_3+Y$, where $T_3$ is the weak isospin generator. 
The EW gauge group is embedded by gauging a subgroup 
of $SU(2)_L\times SU(2)_R \times U(1)_V$ and is dynamically broken to $U(1)_Q$ by the techniquark condensate specified above.

One of the composite states of the spectrum is the pseudoscalar associated with the  axial $U(1)$ anomaly of the underlying gauge theory,  the analogous of the $\eta^\prime$ meson in low-energy QCD. We identify it with the pseudoscalar $X$ discussed in the previous section. 
 It is included as a singlet state in a $2\times 2$ unitary matrix $\mathcal{U}$, which also describes the technipions $\bm{\Pi}\equiv(\Pi^1,\Pi^2,\Pi^3)$:
\begin{equation}\label{eq:U}
	\mathcal{U}\; = \; e^{i \Phi /F_\Pi}\;=\; \exp\left[ \frac{i}{F_\Pi}\left(X\,+\, {\bm{\tau}}\cdot {\bm{\Pi}}\right)  \right]\,,
\end{equation}
where $\bm{\tau}\equiv\left(\tau_1,\,\tau_2\,,\tau_3 \right)$ denote the Pauli matrices. The field matrix $\mathcal{U}$ transforms bilinearly under a chiral rotation:
\begin{equation}
	\mathcal{U} \to u_L\,\mathcal{U} \, u_R^\dagger \label{trU}\,,
\end{equation}
with $u_{L/R} \in SU(2)_{L/R}$. The technipion decay constant $F_\Pi$ sets the EW symmetry breaking scale. Indeed, the EW boson masses result as:
$m_W^2 =1/2 \,g_W\, F_\Pi^2$ and $ m_Z^2=1/ 2\sqrt{g_W^2+g_Y^2} F_\Pi^2$,
with $F_\Pi=v=246$ GeV.\footnote{ In Eq. (\ref{eq:U}) we assume the large-$N_T$ relation between the decay constants of the singlet $X$ and the technipions, namely $F_X = F_\Pi \left(1 + \mathcal{O}(1/N_T )\right)$.}
 At LO in the large $N_T$  limit, $m_X \approx 6/N_T\approx 6/d(R)$ TeV \cite{Molinaro:2016oix}, using the Witten-Veneziano relation \cite{Witten:1979vv,Veneziano:1979ec}.

The global axial-vector currents are anomalous and generate couplings of the composite pseudoscalar $X$  with the EW gauge bosons. These interactions are described by the gauged WZW action \cite{Witten:1983tx,Wess:1971yu,Kaymakcalan:1984bz,Kaymakcalan:1983qq,Schechter:1986vs} (see also \cite{Duan:2000dy}) which gives at LO in  the derivative expansion of the theory:
\bea
\begin{split}
 &-\, \frac{i 5C}{F_\Pi}\epsilon_{\mu\nu\rho\sigma}{\rm Tr}\big[\Phi\big(\partial^\mu A_L^\nu\partial^\rho A_L^\sigma+\partial^\mu A_R^\nu\partial^\rho A_R^\sigma
 \,+\,\partial^\mu\left(A_L^\nu+A_R^\nu \right)\partial^\rho\left(A_L^\sigma+A_R^\sigma \right) \big)\big] \\
& +\, \frac{5C}{F_\Pi^3}\epsilon_{\mu\nu\rho\sigma}{\rm Tr}\big[\partial^\mu\Phi\partial^\nu\Phi\partial^\rho\Phi\left(A_L^\sigma+A_R^\sigma\right)\big]+\ldots\,,
\label{WZW}
\end{split}
\eea
with  $C=-i d(R)/(240\,\pi^2)$ and
\begin{eqnarray}
 A^\mu_L &=&   g_Y \left(Q-\frac 12 \tau_3 \right)\,B^\mu\,+\, \frac 12\, g_W\, \bm{\tau}\cdot\bm{W}^\mu\,, \quad\quad  A^\mu_R \;=\;  g_Y \,Q\,  B^\mu\,. \label{AR} 
\end{eqnarray}
 The latter transform under the EW gauge group as 
\begin{eqnarray}
	A^\mu_L & \to & u_L \,A^\mu_L \,u_L^\dagger\,-\,i\,\partial_\mu u_L\, u_L^\dagger\,,\quad\quad	 A^\mu_R \; \to \; u_R A^\mu_R\, u_R^\dagger\,-\,i\,\partial_\mu u_R\, u_R^\dagger\,.
\end{eqnarray}
where $u_L\in SU(2)_W$ and $u_R\equiv \exp(i\, \theta(x) \,\tau_3/2)$.

The $X$ production mechanisms and decays at the LHC are determined by these topological terms. By matching the interactions in Eq.~(\ref{WZW}) with the effective Lagrangian in Eq.~(\ref{eq:L-eff}) we obtain:
\begin{align}
\begin{split}
	c_{AA} =& \left(1+y^2\right)\, e^2\,\frac{d(R)}{8\,\pi^2}\,, \quad\quad \quad\quad c_{AZ} \;=\; \frac{1-2(1+y^2)s_W^2}{2\,c_W\,s_W}\, e^2\,\frac{d(R)}{8\,\pi^2}\,,\\
	c_{ZZ} =& e^2\,\frac{1-3s_W^2+3(1+y^2)s_W^4}{3\,c_W^2\,s_W^2}\frac{d(R)}{8\,\pi^2}\,,\quad\quad c_{WW} \;=\; e^2\,\frac{1}{s_W^2}\frac{d(R)}{24\,\pi^2}\,.       \label{effc}    
\end{split}
\end{align}
Notice that three-body decay processes $X\to \Pi~\Pi~V$, $V=\gamma,Z,W^{\pm}$, which arise from the second line in Eq.~(\ref{WZW}), dominate over the two-body decays into EW bosons for large masses, i.e. $m_X \gtrsim 1.5$ TeV \cite{Molinaro:2016oix}. 
  
 The gauged WZW term is responsible for the production of $X$ at the LHC and its decays into EW gauge bosons. In this case the dominant production mechanism is via photon fusion. We can use our results for the constraints on the  effective coupling $c_{AA}$ derived in section \ref{subsec:photon-fusion} (see Fig. \ref{fig:aaF})  to obtain information on the underlying theory. Notice that in this model the branching ratio in diphoton depends on  $m_X$ and $y$ and has an upper limit   $\text{BR}(X\to \gamma \gamma)\lesssim 0.7$. The expressions of the decay rates in terms of the effective coefficients in Eq.~(\ref{eq:L-eff}) are reported in \ref{appendice}.
  Taking as benchmark $d(R)=6$ and $m_X=1$ TeV we have that values of the hypercharge parameter $y\geq 1.5$ are excluded at 95\% C.L.,  while with an integrated luminosity of 50 (100) [300] fb$^{-1}$ it is possible to test at LHC-13  values of $y \gtrsim 1.2 ~(1.0)~[0.86]$. 
 
 We now consider the possibility that the new pseudoscalar resonance is directly coupled to SM fermions. This scenario can be realized by an extended gauge dynamics which generates a coupling of $X$ to the top quark as in the effective Lagrangian in Eq. (\ref{eq:L-eff}).\footnote{We refer to \cite{Molinaro:2016oix} for a detailed discussion of the $X$ interactions with SM fermions.} In this case the dominant production mechanism of the new pseudoscalar is via top-mediated gluon fusion. Taking into account the results derived in section \ref{sec:GF} (see Fig. \ref{fig:ggF}) we find that for $d(R)=6$ and $m_X=1$ TeV values of the hypercharge parameter $y\geq 3$ are excluded at 95\% C.L.,  while with an integrated luminosity of 50 (100) [300] fb$^{-1}$ it is possible to test at LHC-13  values of $y \gtrsim 2.2 ~(1.9)~[1.6]$.

 \section{Summary}
\label{summary} 

We have reviewed the present constraints on diphoton resonances at the LHC and indicated the perspectives of detecting a new pseudoscalar state $X$ in the diphoton channel. 
Compared to the previous study in \cite{Molinaro:2016oix}, here we have adopted the new set of photon PDF, LUXqed \cite{Manohar:2016nzj}.

  We have analyzed two minimal scenarios for the $X$ production mechanisms, namely $i)$ photon fusion and $ii)$ top-mediated gluon fusion. We have used an effective description, Eq. (\ref{eq:L-eff}), to determine the current exclusion limit and the future reach on the relevant Wilson coefficient $c_{AA}$ as a function of the $X$ mass, $m_X$.
 Our final results are presented in Fig. \ref{fig:aaF} for  scenario $i)$ and in Fig. \ref{fig:ggF} for scenario $ii)$.  In the first case we find that it is possible to probe the  diphoton resonance up to masses of $\sim 4$ TeV and  $c_{AA}$ as small as $\sim0.008$, with an integrated luminosity of 300 fb$^{-1}$. In the second case the reach in the diphoton channel with 300 fb$^{-1}$ of data  extends up to masses $m_X\lesssim 2$ TeV and  values of $c_{AA}\gtrsim 0.014$.
  
 We have applied this analysis to a specific scenario of dynamical EW symmetry breaking, which naturally includes a new pseudoscalar composite state in the TeV mass range. This state is associated with a global anomalous symmetry of the composite sector, in analogy with the $\eta^\prime$  meson in low-energy QCD.
  Its anomalous couplings with the EW gauge bosons are univocally predicted by the gauged WZW action of the composite theory. They depend on the fundamental parameters of the underlying gauge dynamics, which can be accessed via this analysis of the diphoton channel. 
  
  In general, topological interactions, which are predicted in theories of composite (Goldstone) Higgs, lead to interesting phenomenology both at the LHC and at future proton-proton colliders (see, e.g., \cite{Molinaro:2017mwb}), which may reveal the fundamental mechanism of EW symmetry breaking.
  
\section*{Acknowledgments}
The CP$^{3}$-Origins center is partially funded by the Danish National Research Foundation, grant number DNRF90.

\newpage
\appendix
\section{Decay rates}\label{appendice}

From the effective Lagrangian in Eq.~(\ref{eq:L-eff}), the partial decay rates of $X$ read:
 \begin{eqnarray}
 	&&\Gamma(X \to g g) =  \frac{m_X^3}{8\,\pi}\, \frac{\left|c_{GG}\right|^2}{v^2} \label{gammaGG} \\
	&&\Gamma(X \to \gamma\gamma) \,= \, \frac{m_{X}^3}{64\,\pi}\,\frac{c_{AA}^2}{v^2}\,, \label{eq:rateGammaGamma}\\
	&&\Gamma(X \to \gamma Z) \,= \, \frac{m_{X}^3}{32\,\pi}\,\frac{c_{AZ}^2}{v^2}\left(1\,-\,\frac{m_Z^2}{m_X^2}\right)^3 \, , \\
	&&\Gamma(X \to Z Z) \,= \, \frac{m_{X}^3}{64\,\pi}\,\frac{c_{ZZ}^2}{v^2}\left(1\,-\,\frac{4\,m_Z^2}{m_X^2}\right)^{3/2}\,  ,\\
	&&\Gamma(X \to W^+W^-) \,= \, \frac{m_{X}^3}{32\,\pi}\,\frac{c_{WW}^2}{v^2}\left(1\,-\,\frac{4\,m_W^2}{m_X^2}\right)^{3/2}\, \label{eq:rateWW} ,\\
	&&\Gamma(X \to t\bar{t}) \,= \, y^2_t \, \frac{3\,m_X}{8 \pi}\frac{m_t^2}{v^2}\, \sqrt{1-\frac{4\,m_t^2}{m_X^2}}\label{gammattbar}\,.
 \end{eqnarray}
For the model described in section~\ref{sec:model}, the effective couplings $c_{VV}$ are given in Eq.~\eqref{effc}.  In this case, taking the limit $m_{\Pi^\pm}\approx m_{\Pi^0}\equiv m_\Pi$, the three-body  partial decay rate of $X$  is
\begin{eqnarray}
\begin{split}
 	\Gamma(X \to \Pi~\Pi~V)&= \frac{m_X^3}{122880\,\pi^3}\,\frac{m_X^4}{F_\Pi^6}\Bigg[\sqrt{1-4u^2}\Big(1-2u^2\left(14+47u^2-80u^4+60u^6\right)\Big)\\
					& +\,240 u^4\left(-1+2u^2-3u^4+2u^6\right)\ln\left(\frac{2 u}{1+\sqrt{1-4u^2}}\right)\Bigg]\,c_{\Pi\Pi V}^2\,, \label{3dec}
\end{split}					
 \end{eqnarray}
where $u\equiv m_\Pi/m_X$ and
\begin{eqnarray}
\begin{split}
	c_{\Pi^+\Pi^- \gamma}  & =  e\,\frac{d(R)}{12\,\pi^2}\, ,\\
		c_{\Pi^+\Pi^-  Z} & =  \frac{1-2s_W^2}{2c_W s_W}\,c_{\Pi^+\Pi^-  \gamma}\,,\\
		c_{\Pi^\pm\Pi^0 W^\pm} & = \pm \,
 \frac{1}{2 s_W}\,c_{\Pi^+\Pi^- \gamma}\,.
 \end{split}
\end{eqnarray}


\begin{thebibliography}{999}                                                                                               

%\cite{Aad:2012tfa}
\bibitem{Aad:2012tfa} 
  G.~Aad {\it et al.} [ATLAS Collaboration],
  %``Observation of a new particle in the search for the Standard Model Higgs boson with the ATLAS detector at the LHC,''
  Phys.\ Lett.\ B {\bf 716}, 1 (2012)
  doi:10.1016/j.physletb.2012.08.020
  [arXiv:1207.7214 [hep-ex]].
  %%CITATION = doi:10.1016/j.physletb.2012.08.020;%%
  %7384 citations counted in INSPIRE as of 03 Jul 2017


%\cite{Chatrchyan:2012xdj}
\bibitem{Chatrchyan:2012xdj} 
  S.~Chatrchyan {\it et al.} [CMS Collaboration],
  %``Observation of a new boson at a mass of 125 GeV with the CMS experiment at the LHC,''
  Phys.\ Lett.\ B {\bf 716}, 30 (2012)
  doi:10.1016/j.physletb.2012.08.021
  [arXiv:1207.7235 [hep-ex]].
  %%CITATION = doi:10.1016/j.physletb.2012.08.021;%%
  %7243 citations counted in INSPIRE as of 03 Jul 2017


%\cite{ATLAS:2016eeo}
\bibitem{ATLAS:2016eeo} 
  The ATLAS collaboration [ATLAS Collaboration],
  %``Search for scalar diphoton resonances with 15.4~fb$^{-1}$ of data collected at $\sqrt{s}$=13 TeV in 2015 and 2016 with the ATLAS detector,''
  ATLAS-CONF-2016-059.
  %%CITATION = ATLAS-CONF-2016-059;%%
  %83 citations counted in INSPIRE as of 03 Jul 2017


%\cite{Khachatryan:2016yec}
\bibitem{Khachatryan:2016yec} 
  V.~Khachatryan {\it et al.} [CMS Collaboration],
  %``Search for high-mass diphoton resonances in proton?proton collisions at 13 TeV and combination with 8 TeV search,''
  Phys.\ Lett.\ B {\bf 767}, 147 (2017)
  doi:10.1016/j.physletb.2017.01.027
  [arXiv:1609.02507 [hep-ex]].
  %%CITATION = doi:10.1016/j.physletb.2017.01.027;%%
  %38 citations counted in INSPIRE as of 03 Jul 2017


%\cite{DiVecchia:1980xq}
\bibitem{DiVecchia:1980xq} 
  P.~Di Vecchia and G.~Veneziano,
  %``Minimal Composite Higgs Systems,''
  Phys.\ Lett.\  {\bf 95B}, 247 (1980).
  doi:10.1016/0370-2693(80)90480-3
  %%CITATION = doi:10.1016/0370-2693(80)90480-3;%%
  %21 citations counted in INSPIRE as of 03 Jul 2017


%\cite{Molinaro:2015cwg}
\bibitem{Molinaro:2015cwg} 
  E.~Molinaro, F.~Sannino and N.~Vignaroli,
  %``Minimal Composite Dynamics versus Axion Origin of the Diphoton excess,''
  Mod.\ Phys.\ Lett.\ A {\bf 31}, no. 26, 1650155 (2016)
  doi:10.1142/S0217732316501558
  [arXiv:1512.05334 [hep-ph]].
  %%CITATION = doi:10.1142/S0217732316501558;%%
  %198 citations counted in INSPIRE as of 03 Jul 2017


%\cite{Molinaro:2016oix}
\bibitem{Molinaro:2016oix} 
  E.~Molinaro, F.~Sannino and N.~Vignaroli,
  %``Collider Tests of (Composite) Diphoton Resonances,''
  Nucl.\ Phys.\ B {\bf 911}, 106 (2016)
  doi:10.1016/j.nuclphysb.2016.07.032
  [arXiv:1602.07574 [hep-ph]].
  %%CITATION = doi:10.1016/j.nuclphysb.2016.07.032;%%
  %22 citations counted in INSPIRE as of 03 Jul 2017


%\cite{Tandean:1995ci}
\bibitem{Tandean:1995ci} 
  J.~Tandean,
  %``Observing the technieta at a photon linear collider,''
  Phys.\ Rev.\ D {\bf 52}, 1398 (1995)
  doi:10.1103/PhysRevD.52.1398
  [hep-ph/9505256].
  %%CITATION = doi:10.1103/PhysRevD.52.1398;%%
  %10 citations counted in INSPIRE as of 03 Jul 2017


%\cite{Gripaios:2009pe}
\bibitem{Gripaios:2009pe} 
  B.~Gripaios, A.~Pomarol, F.~Riva and J.~Serra,
  %``Beyond the Minimal Composite Higgs Model,''
  JHEP {\bf 0904}, 070 (2009)
  doi:10.1088/1126-6708/2009/04/070
  [arXiv:0902.1483 [hep-ph]].
  %%CITATION = doi:10.1088/1126-6708/2009/04/070;%%
  %178 citations counted in INSPIRE as of 03 Jul 2017


%\cite{Galloway:2010bp}
\bibitem{Galloway:2010bp} 
  J.~Galloway, J.~A.~Evans, M.~A.~Luty and R.~A.~Tacchi,
  %``Minimal Conformal Technicolor and Precision Electroweak Tests,''
  JHEP {\bf 1010}, 086 (2010)
  doi:10.1007/JHEP10(2010)086
  [arXiv:1001.1361 [hep-ph]].
  %%CITATION = doi:10.1007/JHEP10(2010)086;%%
  %104 citations counted in INSPIRE as of 03 Jul 2017


%\cite{Bellazzini:2015nxw}
\bibitem{Bellazzini:2015nxw} 
  B.~Bellazzini, R.~Franceschini, F.~Sala and J.~Serra,
  %``Goldstones in Diphotons,''
  JHEP {\bf 1604}, 072 (2016)
  doi:10.1007/JHEP04(2016)072
  [arXiv:1512.05330 [hep-ph]].
  %%CITATION = doi:10.1007/JHEP04(2016)072;%%
  %193 citations counted in INSPIRE as of 03 Jul 2017

%\cite{Xue:2016txt}
\bibitem{Xue:2016txt} 
  S.~S.~Xue,
  %``An effective strong-coupling theory of composite particles in UV-domain,''
  JHEP {\bf 1705}, 146 (2017)
  doi:10.1007/JHEP05(2017)146
  [arXiv:1601.06845 [hep-ph]].
  %%CITATION = doi:10.1007/JHEP05(2017)146;%%
  %3 citations counted in INSPIRE as of 03 Aug 2017


%\cite{Mimasu:2014nea}
\bibitem{Mimasu:2014nea} 
  K.~Mimasu and V.~Sanz,
  %``ALPs at Colliders,''
  JHEP {\bf 1506}, 173 (2015)
  doi:10.1007/JHEP06(2015)173
  [arXiv:1409.4792 [hep-ph]].
  %%CITATION = doi:10.1007/JHEP06(2015)173;%%
  %17 citations counted in INSPIRE as of 03 Jul 2017


%\cite{Randall:1999ee}
\bibitem{Randall:1999ee} 
  L.~Randall and R.~Sundrum,
  %``A Large mass hierarchy from a small extra dimension,''
  Phys.\ Rev.\ Lett.\  {\bf 83}, 3370 (1999)
  doi:10.1103/PhysRevLett.83.3370
  [hep-ph/9905221].
  %%CITATION = doi:10.1103/PhysRevLett.83.3370;%%
  %7561 citations counted in INSPIRE as of 03 Jul 2017


%\cite{Goldberger:1999uk}
\bibitem{Goldberger:1999uk} 
  W.~D.~Goldberger and M.~B.~Wise,
  %``Modulus stabilization with bulk fields,''
  Phys.\ Rev.\ Lett.\  {\bf 83}, 4922 (1999)
  doi:10.1103/PhysRevLett.83.4922
  [hep-ph/9907447].
  %%CITATION = doi:10.1103/PhysRevLett.83.4922;%%
  %1161 citations counted in INSPIRE as of 03 Jul 2017


%\cite{Sundrum:2003yt}
\bibitem{Sundrum:2003yt} 
  R.~Sundrum,
  %``Gravity's scalar cousin,''
  hep-th/0312212.
  %%CITATION = HEP-TH/0312212;%%
  %21 citations counted in INSPIRE as of 03 Jul 2017


%\cite{Coradeschi:2013gda}
\bibitem{Coradeschi:2013gda} 
  F.~Coradeschi, P.~Lodone, D.~Pappadopulo, R.~Rattazzi and L.~Vitale,
  %``A naturally light dilaton,''
  JHEP {\bf 1311}, 057 (2013)
  doi:10.1007/JHEP11(2013)057
  [arXiv:1306.4601 [hep-th]].
  %%CITATION = doi:10.1007/JHEP11(2013)057;%%
  %37 citations counted in INSPIRE as of 03 Jul 2017


%\cite{Bellazzini:2013fga}
\bibitem{Bellazzini:2013fga} 
  B.~Bellazzini, C.~Csaki, J.~Hubisz, J.~Serra and J.~Terning,
  %``A Naturally Light Dilaton and a Small Cosmological Constant,''
  Eur.\ Phys.\ J.\ C {\bf 74}, 2790 (2014)
  doi:10.1140/epjc/s10052-014-2790-x
  [arXiv:1305.3919 [hep-th]].
  %%CITATION = doi:10.1140/epjc/s10052-014-2790-x;%%
  %42 citations counted in INSPIRE as of 03 Jul 2017


%\cite{Megias:2014iwa}
\bibitem{Megias:2014iwa} 
  E.~Megias and O.~Pujolas,
  %``Naturally light dilatons from nearly marginal deformations,''
  JHEP {\bf 1408}, 081 (2014)
  doi:10.1007/JHEP08(2014)081
  [arXiv:1401.4998 [hep-th]].
  %%CITATION = doi:10.1007/JHEP08(2014)081;%%
  %21 citations counted in INSPIRE as of 03 Jul 2017


%\cite{Vecchi:2010gj}
\bibitem{Vecchi:2010gj} 
  L.~Vecchi,
  %``Phenomenology of a light scalar: the dilaton,''
  Phys.\ Rev.\ D {\bf 82}, 076009 (2010)
  doi:10.1103/PhysRevD.82.076009
  [arXiv:1002.1721 [hep-ph]].
  %%CITATION = doi:10.1103/PhysRevD.82.076009;%%
  %67 citations counted in INSPIRE as of 03 Jul 2017

%\cite{Ahmed:2015uqt}
\bibitem{Ahmed:2015uqt} 
  A.~Ahmed, B.~M.~Dillon, B.~Grzadkowski, J.~F.~Gunion and Y.~Jiang,
  %``Implications of the absence of high-mass radion signals,''
  Phys.\ Rev.\ D {\bf 95}, no. 9, 095019 (2017)
  doi:10.1103/PhysRevD.95.095019
  [arXiv:1512.05771 [hep-ph]].
  %%CITATION = doi:10.1103/PhysRevD.95.095019;%%
  %169 citations counted in INSPIRE as of 10 Jul 2017
  
  
 

%\cite{Gopalakrishna:2017zku}
\bibitem{Gopalakrishna:2017zku} 
  S.~Gopalakrishna and T.~S.~Mukherjee,
  %``Gauge Singlet Vector-like Fermion Dark Matter, LHC Diphoton Rate and Direct Detection,''
  arXiv:1702.04000 [hep-ph].
  %%CITATION = ARXIV:1702.04000;%%


%\cite{Bian:2017jpt}
\bibitem{Bian:2017jpt} 
  L.~Bian, N.~Chen and Y.~Zhang,
  %``CP violation effects in the diphoton spectrum of heavy scalars,''
  arXiv:1706.09425 [hep-ph].
  %%CITATION = ARXIV:1706.09425;%%


%\cite{Bellazzini:2017neg}
\bibitem{Bellazzini:2017neg} 
  B.~Bellazzini, A.~Mariotti, D.~Redigolo, F.~Sala and J.~Serra,
  %``R-axion at colliders,''
  arXiv:1702.02152 [hep-ph].
  %%CITATION = ARXIV:1702.02152;%%


%\cite{Cao:2016uwt}
\bibitem{Cao:2016uwt} 
  J.~Cao, X.~Guo, Y.~He, P.~Wu and Y.~Zhang,
  %``Diphoton signal of the light Higgs boson in natural NMSSM,''
  Phys.\ Rev.\ D {\bf 95}, no. 11, 116001 (2017)
  doi:10.1103/PhysRevD.95.116001
  [arXiv:1612.08522 [hep-ph]].
  %%CITATION = doi:10.1103/PhysRevD.95.116001;%%
  %2 citations counted in INSPIRE as of 03 Jul 2017


%\cite{Allanach:2017qbs}
\bibitem{Allanach:2017qbs} 
  B.~C.~Allanach, D.~Bhatia and A.~M.~Iyer,
  %``Dissecting Multi-Photon Resonances at the Large Hadron Collider,''
  arXiv:1706.09039 [hep-ph].
  %%CITATION = ARXIV:1706.09039;%%


%\cite{Witten:1983tx}
\bibitem{Witten:1983tx} 
  E.~Witten,
  %``Current Algebra, Baryons, and Quark Confinement,''
  Nucl.\ Phys.\ B {\bf 223}, 433 (1983).
  doi:10.1016/0550-3213(83)90064-0
  %%CITATION = doi:10.1016/0550-3213(83)90064-0;%%
  %1254 citations counted in INSPIRE as of 03 Jul 2017


%\cite{Wess:1971yu}
\bibitem{Wess:1971yu} 
  J.~Wess and B.~Zumino,
  %``Consequences of anomalous Ward identities,''
  Phys.\ Lett.\  {\bf 37B}, 95 (1971).
  doi:10.1016/0370-2693(71)90582-X
  %%CITATION = doi:10.1016/0370-2693(71)90582-X;%%
  %2420 citations counted in INSPIRE as of 03 Jul 2017


%\cite{Kaymakcalan:1984bz}
\bibitem{Kaymakcalan:1984bz} 
  O.~Kaymakcalan and J.~Schechter,
  %``Chiral Lagrangian of Pseudoscalars and Vectors,''
  Phys.\ Rev.\ D {\bf 31}, 1109 (1985).
  doi:10.1103/PhysRevD.31.1109
  %%CITATION = doi:10.1103/PhysRevD.31.1109;%%
  %196 citations counted in INSPIRE as of 03 Jul 2017


%\cite{Kaymakcalan:1983qq}
\bibitem{Kaymakcalan:1983qq} 
  O.~Kaymakcalan, S.~Rajeev and J.~Schechter,
  %``Nonabelian Anomaly and Vector Meson Decays,''
  Phys.\ Rev.\ D {\bf 30}, 594 (1984).
  doi:10.1103/PhysRevD.30.594
  %%CITATION = doi:10.1103/PhysRevD.30.594;%%
  %414 citations counted in INSPIRE as of 03 Jul 2017


%\cite{Schechter:1986vs}
\bibitem{Schechter:1986vs} 
  J.~Schechter,
  %``Electromagnetism in a Gauged Chiral Model,''
  Phys.\ Rev.\ D {\bf 34}, 868 (1986).
  doi:10.1103/PhysRevD.34.868
  %%CITATION = doi:10.1103/PhysRevD.34.868;%%
  %64 citations counted in INSPIRE as of 03 Jul 2017


%\cite{Steinberger:1949wx}
\bibitem{Steinberger:1949wx} 
  J.~Steinberger,
  %``On the Use of subtraction fields and the lifetimes of some types of meson decay,''
  Phys.\ Rev.\  {\bf 76}, 1180 (1949).
  doi:10.1103/PhysRev.76.1180
  %%CITATION = doi:10.1103/PhysRev.76.1180;%%
  %276 citations counted in INSPIRE as of 03 Jul 2017


%\cite{Fichet:2013gsa}
\bibitem{Fichet:2013gsa} 
  S.~Fichet, G.~von Gersdorff, O.~Kepka, B.~Lenzi, C.~Royon and M.~Saimpert,
  %``Probing new physics in diphoton production with proton tagging at the Large Hadron Collider,''
  Phys.\ Rev.\ D {\bf 89}, 114004 (2014)
  doi:10.1103/PhysRevD.89.114004
  [arXiv:1312.5153 [hep-ph]].
  %%CITATION = doi:10.1103/PhysRevD.89.114004;%%
  %43 citations counted in INSPIRE as of 03 Jul 2017


%\cite{Fichet:2014uka}
\bibitem{Fichet:2014uka} 
  S.~Fichet, G.~von Gersdorff, B.~Lenzi, C.~Royon and M.~Saimpert,
  %``Light-by-light scattering with intact protons at the LHC: from Standard Model to New Physics,''
  JHEP {\bf 1502}, 165 (2015)
  doi:10.1007/JHEP02(2015)165
  [arXiv:1411.6629 [hep-ph]].
  %%CITATION = doi:10.1007/JHEP02(2015)165;%%
  %43 citations counted in INSPIRE as of 03 Jul 2017


%\cite{Harland-Lang:2016kog}
\bibitem{Harland-Lang:2016kog} 
  L.~A.~Harland-Lang, V.~A.~Khoze and M.~G.~Ryskin,
  %``Photon-initiated processes at high mass,''
  Phys.\ Rev.\ D {\bf 94}, no. 7, 074008 (2016)
  doi:10.1103/PhysRevD.94.074008
  [arXiv:1607.04635 [hep-ph]].
  %%CITATION = doi:10.1103/PhysRevD.94.074008;%%
  %14 citations counted in INSPIRE as of 01 Sep 2017

%\cite{Alwall:2014hca}
\bibitem{Alwall:2014hca} 
  J.~Alwall {\it et al.},
  %``The automated computation of tree-level and next-to-leading order differential cross sections, and their matching to parton shower simulations,''
  JHEP {\bf 1407}, 079 (2014)
  doi:10.1007/JHEP07(2014)079
  [arXiv:1405.0301 [hep-ph]].
  %%CITATION = doi:10.1007/JHEP07(2014)079;%%
  %1968 citations counted in INSPIRE as of 03 Jul 2017


%\cite{Manohar:2016nzj}
\bibitem{Manohar:2016nzj} 
  A.~Manohar, P.~Nason, G.~P.~Salam and G.~Zanderighi,
  %``How bright is the proton? A precise determination of the photon parton distribution function,''
  Phys.\ Rev.\ Lett.\  {\bf 117}, no. 24, 242002 (2016)
  doi:10.1103/PhysRevLett.117.242002
  [arXiv:1607.04266 [hep-ph]].
  %%CITATION = doi:10.1103/PhysRevLett.117.242002;%%
  %35 citations counted in INSPIRE as of 03 Jul 2017


%\cite{Ball:2013hta}
\bibitem{Ball:2013hta} 
  R.~D.~Ball {\it et al.} [NNPDF Collaboration],
  %``Parton distributions with QED corrections,''
  Nucl.\ Phys.\ B {\bf 877}, 290 (2013)
  doi:10.1016/j.nuclphysb.2013.10.010
  [arXiv:1308.0598 [hep-ph]].
  %%CITATION = doi:10.1016/j.nuclphysb.2013.10.010;%%
  %237 citations counted in INSPIRE as of 03 Jul 2017

%\cite{Martin:2004dh}
\bibitem{Martin:2004dh} 
  A.~D.~Martin, R.~G.~Roberts, W.~J.~Stirling and R.~S.~Thorne,
  %``Parton distributions incorporating QED contributions,''
  Eur.\ Phys.\ J.\ C {\bf 39}, 155 (2005)
  doi:10.1140/epjc/s2004-02088-7
  [hep-ph/0411040].
  %%CITATION = doi:10.1140/epjc/s2004-02088-7;%%
  %374 citations counted in INSPIRE as of 03 Jul 2017


%\cite{Schmidt:2015zda}
\bibitem{Schmidt:2015zda} 
  C.~Schmidt, J.~Pumplin, D.~Stump and C.~P.~Yuan,
  %``CT14QED parton distribution functions from isolated photon production in deep inelastic scattering,''
  Phys.\ Rev.\ D {\bf 93}, no. 11, 114015 (2016)
  doi:10.1103/PhysRevD.93.114015
  [arXiv:1509.02905 [hep-ph]].
  %%CITATION = doi:10.1103/PhysRevD.93.114015;%%
  %46 citations counted in INSPIRE as of 03 Jul 2017


%\cite{Demartin:2014fia}
\bibitem{Demartin:2014fia} 
  F.~Demartin, F.~Maltoni, K.~Mawatari, B.~Page and M.~Zaro,
  %``Higgs characterisation at NLO in QCD: CP properties of the top-quark Yukawa interaction,''
  Eur.\ Phys.\ J.\ C {\bf 74}, no. 9, 3065 (2014)
  doi:10.1140/epjc/s10052-014-3065-2
  [arXiv:1407.5089 [hep-ph]].
  %%CITATION = doi:10.1140/epjc/s10052-014-3065-2;%%
  %56 citations counted in INSPIRE as of 03 Jul 2017


%\cite{Aad:2015mna}
\bibitem{Aad:2015mna} 
  G.~Aad {\it et al.} [ATLAS Collaboration],
  %``Search for high-mass diphoton resonances in $pp$ collisions at $\sqrt{s}=8$ TeV with the ATLAS detector,''
  Phys.\ Rev.\ D {\bf 92}, no. 3, 032004 (2015)
  doi:10.1103/PhysRevD.92.032004
  [arXiv:1504.05511 [hep-ex]].
  %%CITATION = doi:10.1103/PhysRevD.92.032004;%%
  %154 citations counted in INSPIRE as of 03 Jul 2017


%\cite{Khachatryan:2015qba}
\bibitem{Khachatryan:2015qba} 
  V.~Khachatryan {\it et al.} [CMS Collaboration],
  %``Search for diphoton resonances in the mass range from 150 to 850 GeV in pp collisions at $\sqrt{s} =$ 8 TeV,''
  Phys.\ Lett.\ B {\bf 750}, 494 (2015)
  doi:10.1016/j.physletb.2015.09.062
  [arXiv:1506.02301 [hep-ex]].
  %%CITATION = doi:10.1016/j.physletb.2015.09.062;%%
  %129 citations counted in INSPIRE as of 03 Jul 2017


%\cite{Aaboud:2016tru}
\bibitem{Aaboud:2016tru} 
  M.~Aaboud {\it et al.} [ATLAS Collaboration],
  %``Search for resonances in diphoton events at $\sqrt{s}$=13 TeV with the ATLAS detector,''
  JHEP {\bf 1609}, 001 (2016)
  doi:10.1007/JHEP09(2016)001
  [arXiv:1606.03833 [hep-ex]].
  %%CITATION = doi:10.1007/JHEP09(2016)001;%%
  %109 citations counted in INSPIRE as of 03 Jul 2017


%\cite{Sjostrand:2006za}
\bibitem{Sjostrand:2006za} 
  T.~Sjostrand, S.~Mrenna and P.~Z.~Skands,
  %``PYTHIA 6.4 Physics and Manual,''
  JHEP {\bf 0605}, 026 (2006)
  doi:10.1088/1126-6708/2006/05/026
  [hep-ph/0603175].
  %%CITATION = doi:10.1088/1126-6708/2006/05/026;%%
  %8715 citations counted in INSPIRE as of 03 Jul 2017


%\cite{deFavereau:2013fsa}
\bibitem{deFavereau:2013fsa} 
  J.~de Favereau {\it et al.} [DELPHES 3 Collaboration],
  %``DELPHES 3, A modular framework for fast simulation of a generic collider experiment,''
  JHEP {\bf 1402}, 057 (2014)
  doi:10.1007/JHEP02(2014)057
  [arXiv:1307.6346 [hep-ex]].
  %%CITATION = doi:10.1007/JHEP02(2014)057;%%
  %716 citations counted in INSPIRE as of 03 Jul 2017


%\cite{Witten:1982fp}
\bibitem{Witten:1982fp} 
  E.~Witten,
  %``An SU(2) Anomaly,''
  Phys.\ Lett.\  {\bf 117B}, 324 (1982).
  doi:10.1016/0370-2693(82)90728-6
  %%CITATION = doi:10.1016/0370-2693(82)90728-6;%%
  %782 citations counted in INSPIRE as of 03 Jul 2017


%\cite{Witten:1979vv}
\bibitem{Witten:1979vv} 
  E.~Witten,
  %``Current Algebra Theorems for the U(1) Goldstone Boson,''
  Nucl.\ Phys.\ B {\bf 156}, 269 (1979).
  doi:10.1016/0550-3213(79)90031-2
  %%CITATION = doi:10.1016/0550-3213(79)90031-2;%%
  %1342 citations counted in INSPIRE as of 03 Jul 2017


%\cite{Veneziano:1979ec}
\bibitem{Veneziano:1979ec} 
  G.~Veneziano,
  %``U(1) Without Instantons,''
  Nucl.\ Phys.\ B {\bf 159}, 213 (1979).
  doi:10.1016/0550-3213(79)90332-8
  %%CITATION = doi:10.1016/0550-3213(79)90332-8;%%
  %1182 citations counted in INSPIRE as of 03 Jul 2017


%\cite{Duan:2000dy}
\bibitem{Duan:2000dy} 
  Z.~Y.~Duan, P.~S.~Rodrigues da Silva and F.~Sannino,
  %``Enhanced global symmetry constraints on epsilon terms,''
  Nucl.\ Phys.\ B {\bf 592}, 371 (2001)
  doi:10.1016/S0550-3213(00)00550-2
  [hep-ph/0001303].
  %%CITATION = doi:10.1016/S0550-3213(00)00550-2;%%
  %58 citations counted in INSPIRE as of 03 Jul 2017


%\cite{Molinaro:2017mwb}
\bibitem{Molinaro:2017mwb} 
  E.~Molinaro, F.~Sannino, A.~E.~Thomsen and N.~Vignaroli,
  %``Uncovering new strong dynamics via topological interactions at the 100 TeV collider,''
  arXiv:1706.04037 [hep-ph].
  %%CITATION = ARXIV:1706.04037;%%

\end{thebibliography}
\end{document}